\documentclass[preprint,showpacs,aps,floatfix]{revtex4}
\usepackage{epsf}

\def\beq{\begin{equation}}
\def\eeq{\end{equation}}
\def\bea{\begin{eqnarray}}
\def\eea{\end{eqnarray}} 
\def\beqa{\begin{equation}\begin{array}{l}}
\def\eeqa{\end{array}\end{equation}}
\def\eqlab#1{\label{eq:#1}}
\def\figlab#1{\label{fig:#1}}
\def\tablab#1{\label{tab:#1}}
   
\def\eref#1{(\ref{eq:#1})}
\def\Eqref#1{Eq.~(\ref{eq:#1})}

\def\fref#1{\ref{fig:#1}}
\def\Figref#1{Fig.~\ref{fig:#1}}
\def\tabref#1{\ref{tab:#1}}

\def\half{\mbox{\small{$\frac{1}{2}$}}}

\def\QN{QNs}
\begin{document}

\title{A Statistical Analysis of Hadron Spectrum:\\ 
Quantum Chaos in Hadrons }

\author{Vladimir Pascalutsa}
\email{vlad@phy.ohiou.edu}

\affiliation{The School of Chemistry, Physics 
and Earth Sciences, 
Flinders University,
Bedford Park, 
SA 5042, Australia \\
and\\
Department of Physics and Astronomy, Ohio University,
Athens, OH 45701}\thanks{Present address.}
\date{\today}

\begin{abstract}
The nearest-neighbor mass-spacing 
distribution of the
meson and baryon spectrum (up to 2.5 GeV) is described by
the {\em Wigner surmise} corresponding to the statistics
of the Gaussian {\it orthogonal} ensemble 
of random matrix theory.
This can be viewed as a manifestation of {\it quantum chaos}
in hadrons.
\end{abstract}

% PACS, the Physics and Astronomy
                             % Classification Scheme.
%\keywords{Suggested keywords}%Use showkeys class option if keyword
                              %display desired
\maketitle
\thispagestyle{empty}

Quarks are glued into colorless objects called {\it hadrons},
forming a large and complex {\it hadron spectrum}.
This spectrum is the main source of information
about the quark-gluon interactions in the confinement regime, 
and hadron spectroscopy has been 
receiving a lot of attention both theoretically and experimentally. 

In the study of complex
spectra in general, statistical methods proved to be useful.
One of the early successes of statistical analysis was Wigner's
celebrated discovery of the fact that 
fluctuation properties of complex nuclear
spectra are described by the Random Matrix Theory (RMT) 
\cite{Wigner,Dyson,Porter65,Mehta67,Brody:1981}.
It was later realized
that these properties, referred to as Wigner-Dyson
properties, have a great deal of universality and appear in
spectra of many physical systems, ranging
from quantum dots \cite{Alh00} to lattice gauge 
theories \cite{Ver95,Ver00,Berg:2000be,Bittner:2001aq}. 
  
In this Letter we
show that the experimentally measured hadron spectrum has 
at least some Wigner-Dyson properties as well. 
Once this is established
the link to the underlying dynamics is made
via the Bohigas--Giannoni--Schmit (BGS) 
conjecture \cite{BGS83}, which states 
that the Wigner-Dyson properties are generic to spectra of the
systems with a quantum analog of chaotic dynamics --- the
``quantum chaos'' (see, {\it e.g.}, \cite{Gut90,CaC95}).

Recent theoretical work in lattice QCD \cite{Ver95,Ver00,Berg:2000be} and
quantum-mechanical Yang-Mills models 
\cite{Salasnich:1997cw,Bittner:2001aq} has already  addressed
the statistical properties of QCD spectrum as well as the quantum
chaos aspects. Our findings will provide certain 
empirical support of these results and, hopefully, motivate
further studies in this direction. 

We look at the experimentally measured mass spectrum
of hadrons up to 2.5 GeV taken from the 
Particle Data Group (PDG) Summary Tables~\cite{pdg}. 
More specifically,
we consider $N$, $\Delta$, $\Lambda$, and $\Sigma$
baryons up to $N$(2200), and all the mesons listed in the
Summary Tables up to $f_2(2340)$. In doing so we exclude
the poorly known ``one-star baryons.'' 
We have only verified that they do not have a significant
impact on our results. 

The spectrum can be organized into multiplets, see \Figref{spektr},
characterized by a set of definite quantum numbers (\QN): 
isospin, spin, parity, 
strangeness, baryon number, and, in the case of mesons, 
charge conjugation. 
Using these data we would like to examine the probability 
distribution of spacing (mass-splitting), $S_i=m_{i+1} -m_{i}$,
between nearest-neighbor hadrons 
within one multiplet (same \QN).
We thus do not need to consider channels with only a single
state below 2.5 GeV. 
\begin{figure}[h]
\begin{center}
\epsffile{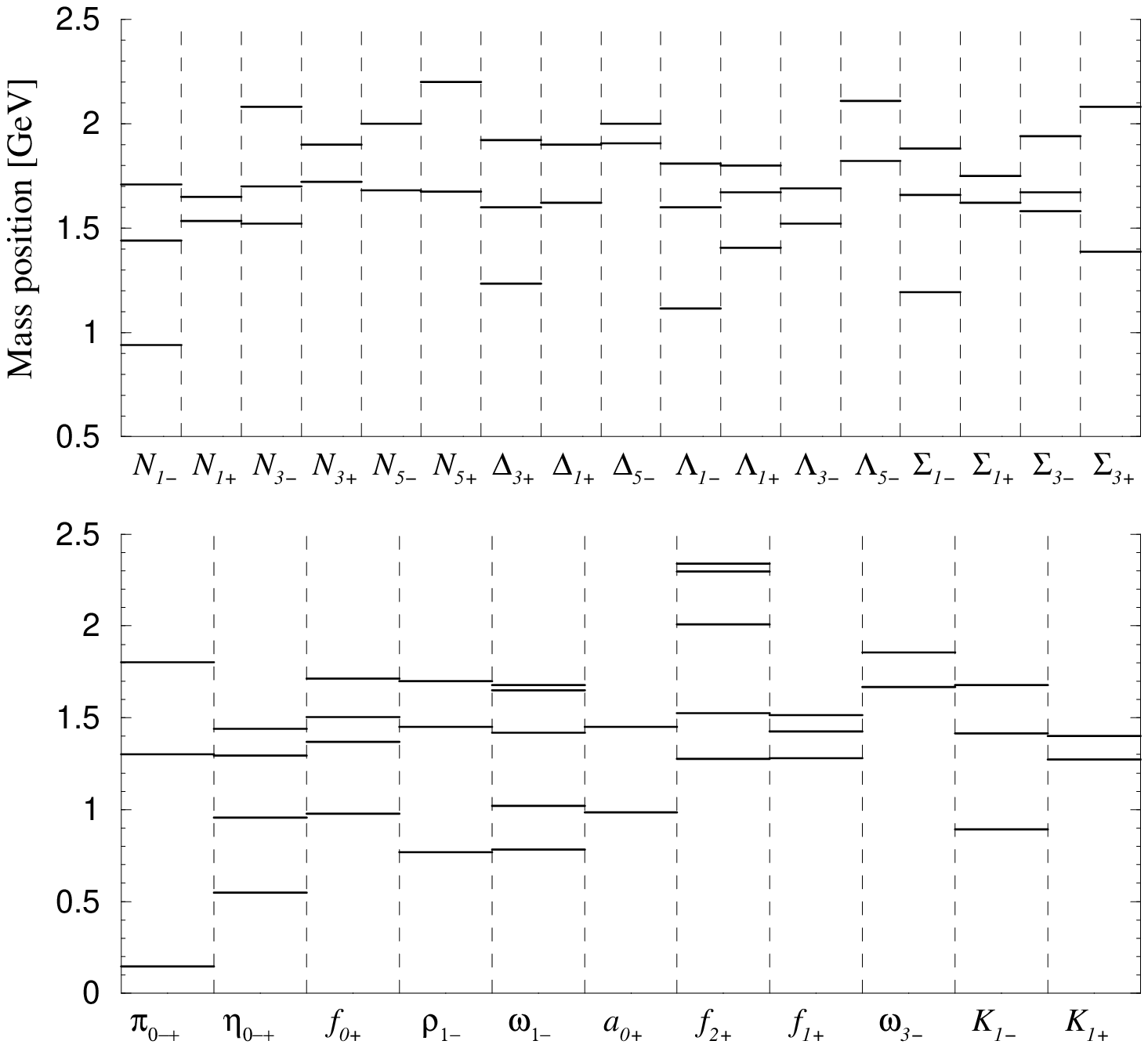}
\end{center}
\caption{\it Light hadron spectrum (excluding 1-star baryons)
in the form of multiplets characterized
by definite quantum numbers.} 
\figlab{spektr}
\end{figure}

Each multiplet provides a number of spacings which are gathered
into a common array of spacings.
We actually will be considering three arrays: $\{S_i\}_{B}$, $\{S_i\}_{M}$,
and $\{S_i\}_{H}=\{ \{S_i\}_{B},\,\{S_i\}_{M}\}$ containing the spacings of, 
respectively, the baryon, the meson, and all
multiplets.
As usual, the mean spacing, $\left<S\right>=(1/N)\sum_{k=1}^{N} S_k$
is scaled out and we deal with dimensionless arrays:
\bea
\{s_i\}_{B}&=& 
\{S_i\}_{B}/\left<S\right>_{B}\,, \nonumber\\
\{s_i\}_{M}&=& \{S_i\}_{M}/\left<S\right>_{M}\,,\\
\{s_i\}_{H} &=& \{S_i\}_{H}/\left<S\right>_{H}=
\{ \{S_i\}_{B}/\left<S\right>_{H},\,\{S_i\}_{M}/\left<S\right>_{H}\}.
\nonumber
\eqlab{norm}
\eea
The mean spacing values $\left<S\right>_{B,M,H}$
are computed in Table~I. 

\begin{table}[h]
\begin{center}
\begin{tabular}{c|c}\hline \hline
Spectrum & $\left<S\right>$, in MeV\\ \hline
Baryon & 289.8 $\pm$ 18.8 \\
Meson & 284.3 $\pm$ 35.8 \\
Hadron & 287.1 $\pm$ 27.3 \\
\hline\hline
\end{tabular}
\end{center}
\caption{Mean mass spacing. The error bars stem from 
the PDG quoted errors.}
\tablab{gparams}
\end{table}

Histograms of the resulting 
spacing distributions, $p(\{s_i\}_{B})$, 
$p(\{s_i\}_{M})$, and $p(\{s_i\}_{H})$
are presented in~\Figref{pofs}. 
The curves in the figure show the Poisson-like distribution
[$p(s)=\exp(-s)$], and the {\it Wigner surmise}
[$p(s)=\frac{\pi}{2}\,s\,\exp(-{\frac{\pi}{4}s^2})$].
They represent two extreme regimes: uncorrelated spectrum
vs.\ a strongly correlated one.
\begin{figure}[h]
\begin{center}
\epsffile{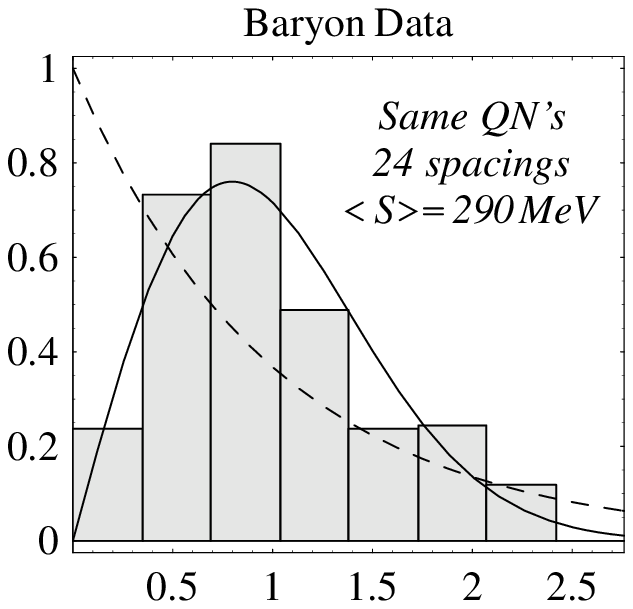}%\\
\epsffile{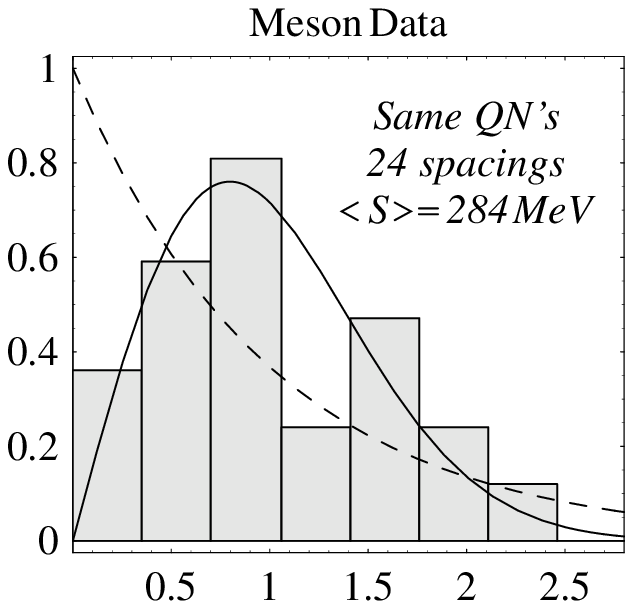}%\\
\epsffile{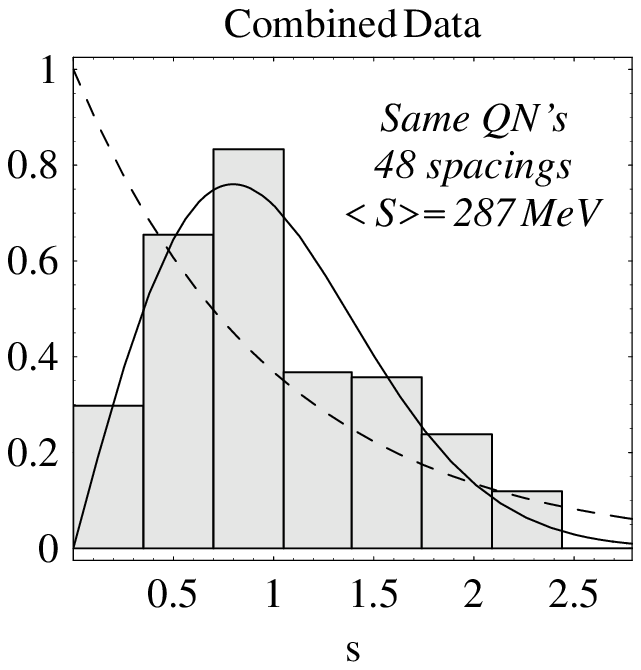}%\\
\end{center}
\caption{Histograms of the nearest-neighbor mass spacing 
distribution for hadron states with same \QN.
Curves represent the Poisson (dashed) and Wigner (solid) 
distributions.} 
\figlab{pofs}
\end{figure}
The figure clearly shows
that the experimental distribution is of the 
Wigner-surmise type. We have checked that
if one does not discriminate the \QN\ (``mixed \QN'')
the spacing distribution is Poissonian, so only 
states with the same \QN\ are correlated.

The histogram plots represent the distributions in
a rather qualitative way because due to the low statistics 
the picture is very sensitive to the choice of the grid. 
Accounting for the error bars makes it even less accurate. 
It is more instructive to consider integrated characteristics, 
such as the {\it moments} of distribution.
For our empirical distribution the $n$-th moment is simply calculated
as
\beq
\left<s^n\right>_{\rm exp} = \frac{1}{N}\sum\limits_{i=1}^N s_i^n \,,
\eeq 
which obviously is not sensitive to the analysis artifacts 
such as the choice of energy grid.

At the same time let us
consider a more general form of the Wigner surmise:
\beq
\eqlab{ws}
p_{\rm W}(s) = A s^\beta\exp(-B s^2)
\eeq
where $\beta$ is a parameter --- the Dyson index; $A$ and $B$ are 
constants fixed
by the normalization conditions: 
$\left<s^0\right>_{\rm W}=1=\left<s^1\right>_{\rm W}$, where,
naturally, $\left<s^n\right>_{\rm W}= \int_0^{\infty}
ds\, s^n\,p_{\rm W}(s)$. 
Depending on the value of $\beta$ expression 
\eref{ws} approximates the spacing 
distributions of the RMT for various types of Gaussian ensembles. 
The most common values are
$\beta =1,\,2$, and $4$ which represent the orthogonal (GOE),
unitary (GUE), and symplectic (GSE) ensembles, 
respectively.

\begin{figure}[h]
\begin{center}
\epsfysize=8cm
\epsffile{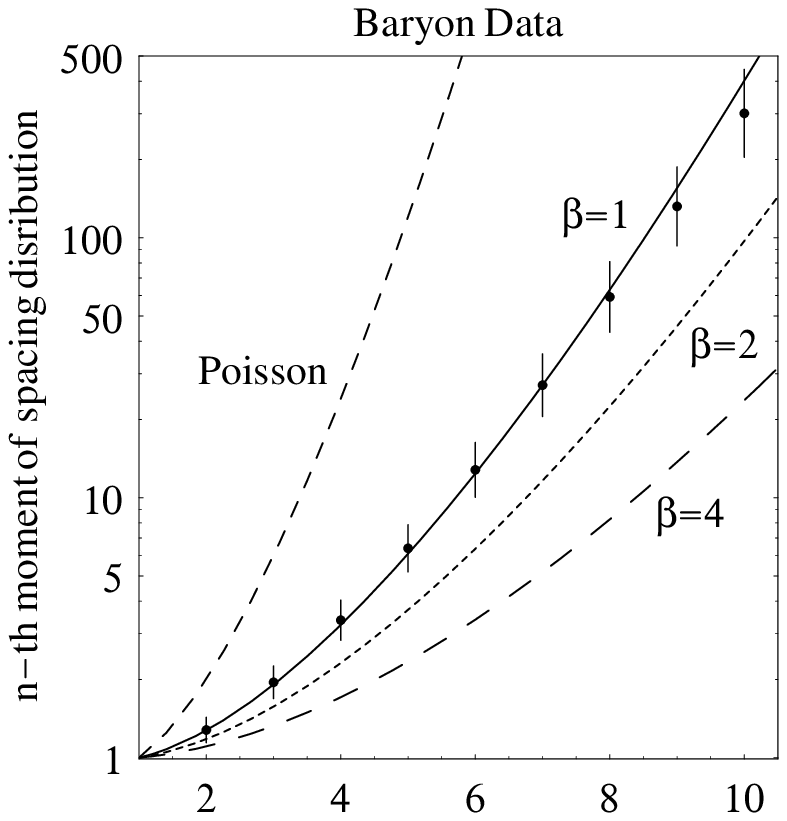}
\epsfysize=8cm
\epsffile{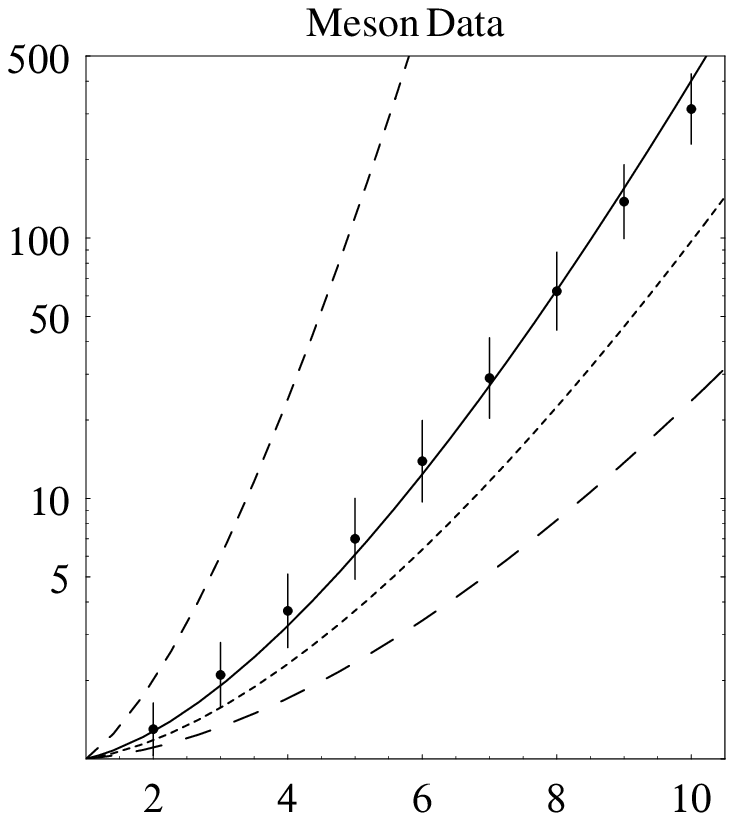}%\\
\epsfysize=8cm
\epsffile{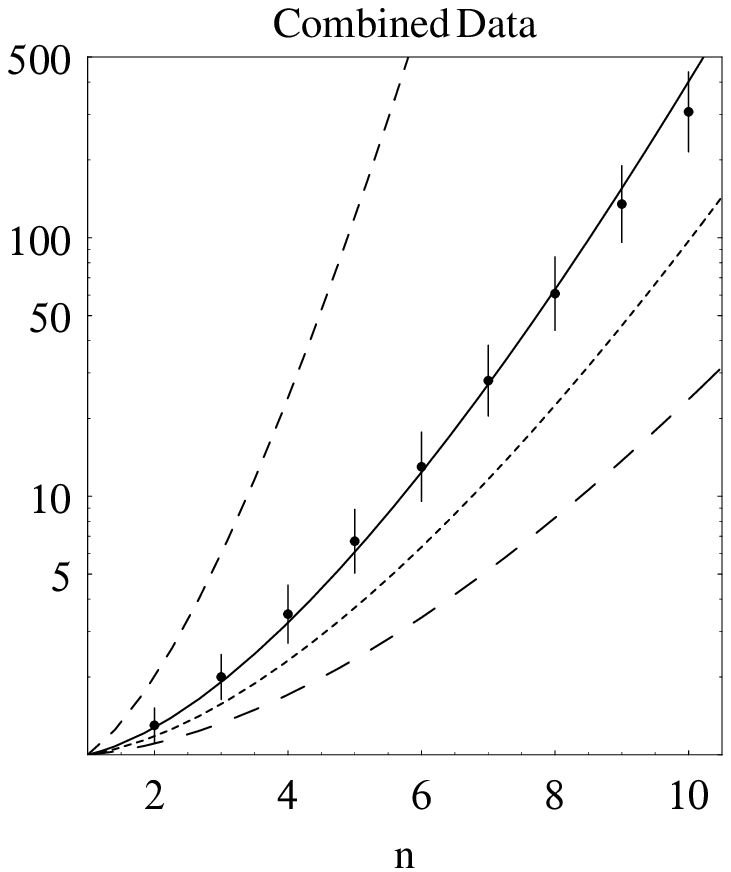}
\end{center}
\caption{ Moments of spacing distributions. Explanations
on the data points are given in the text. Curves represent the
moments of Poisson distribution (dashed), and of the Wigner
surmise for GOE (solid), GUE (short-dashed), and GSE (long-dashed).} 
\figlab{moments}
\end{figure}

Computing the first ten, or so, moments of all the relevant distributions
we obtain Figure~\fref{moments}. The moments of the Poisson and Wigner
distributions (shown by lines) are proportional to 
the Gamma-function [{\em e.g.}, $\left<s^n\right>_P=\Gamma(n+1)$, and
 $\left<s^n\right>_W
=(2/\sqrt{\pi})^n\, \Gamma(\half n+1)$, for $\beta=1$]. Therefore 
they all increase rapidly with $n$ and we have used 
logarithmic scale to plot them.

Data points in the first three panels correspond to the 
empirical distributions shown by the histograms in  
\Figref{pofs}, now with the PDG error bars properly accounted 
for\footnote{We account only for the the quoted mass-position error;
the width is not viewed as one.}.
The Wigner surmise with $\beta=1$ fits these
distributions extremely well: comparing the first ten moments, 
$\chi^2/{\rm point}$ values for the three distributions
are shown in the ``None'' row of Table~\tabref{chisq}.

In general, before combining the level statistics
from two different spectra we would need 
to perform an {\it unfolding} of these spectra. 
Then the combined meson-baryon
array of spacings could be constructed as
$\{s_i\}_{H}'=\{ \{s_i\}_{B},\,\{s_i\}_{M}\}$. However,
because the baryon and meson mean spacing come out so similar
(Table~I)
this construction leads to practically same results as the one in 
\Eqref{norm}, the $\chi^2$ value changes insignificantly, see 
``B vs M'' row of Table~\tabref{chisq}.
One can also perform the unfolding with respect to 
strangeness, which means 
one treats the spectrum of strange baryons and mesons as
different from that of non-strange ones, and unfolds
({\it i.e.}, normalizes to  unit mean spacing) prior combining them. 
In this case the agreement with Wigner surmise has slightly improved
as is seen from the ``S=0 vs $-1$'' row of Table~\tabref{chisq}. 

\begin{table}[h]
\begin{center}
\begin{tabular}{c|c|c|c}\hline \hline
Unfolding & Baryon $\chi^2$ & Meson $\chi^2$ & Hadron $\chi^2$\\ \hline
None & 0.52 & 0.66 & 0.57 \\
B vs M & --- & --- & 0.54 \\
S=0 vs $-1$ & 0.29 & 0.40 & 0.33 \\
\hline\hline
\end{tabular}
\end{center}
\caption{$\chi^2/point$ fit of $\beta=1$ Wigner surmise
to the first ten moments
of the experimental spacing distributions for various unfoldings.}
\tablab{chisq}
\end{table}

To recapitulate, the hadron level-spacing
distribution is remarkably well described by the
Wigner surmise for $\beta=1$. This indicates that
the fluctuation properties of the hadron spectrum 
fall into the GOE universality class, and
hence hadrons exhibit the {\it quantum chaos} phenomenon.
One then should be able to describe 
the statistical properties of hadron spectra
using RMT with random Hamiltonians from GOE
that are characterized by good time-reversal 
and rotational symmetry. 

The hadron-exchange models ({\it e.g.}\cite{PaT00})
should also be able benefit from the RMT methods, in a way
analogous to the stochastic theory of compound-nucleus
reactions \cite{Brody:1981,VWZ85}. However, since unitarity is an
essential ingredient of such models, a study of
hadron width distribution is in order. Widths
of compound nuclei have the Porter--Thomas distribution \cite{Porter65}
that eventually allows one to define the unitary $S$-matrix
completely within the RMT. We anticipate that the same approach will
be applicable in the hadron-exchange reactions once
the hadron-width distribution is determined.

It would be interesting to see how quantum chaos
emerges in hadrons as a result of quark-gluon 
underlying dynamics. Lattice QCD studies in this
direction are underway~\cite{Ver95,Ver00,Berg:2000be}. This problem
could also be addressed in a quark-model framework.  
In particular, does a quark-model Hamiltonians which fits 
the physical spectrum indeed defines a chaotic classical system?

In conclusion, we have examined the nearest-neighbor level-spacing
(mass-splitting) distribution, $p(s)$, of the experimental hadron spectrum.
We focused on the lighter part of the hadron spectrum ($m<2.5 GeV$), 
since it is more reliably known.
The mean level-spacing seems to be the only relevant scale here,
and after it is scaled out the distribution shows universal behavior. 
Unfortunately, the masses and quantum numbers
are well known only
for a few dozen hadron states, so achieving high statistics is 
out of the question.
Nonetheless, the low-statistics analysis we have performed
pinpoints the {\it moments} of the distribution accurately
enough to claim that hadronic $p(s)$ fits the Wigner surmise
with linear level repulsion ($\beta=1$). This indicates
that the spectrum falls into the
GOE universality class of random matrix theory.
Invoking the BGS conjecture, this result is viewed as an
empirical evidence of the ``quantum chaos'' phenomenon in hadrons.

%\medskip
\begin{acknowledgments}
I thank Professor Iraj Afnan
for valuable discussions, and Professors Daniel Phillips and
Jac Verbaarschot for
critical remarks on the manuscript.
The work was supported by the Australian Research Council (ARC)
and in part by DOE under grant DE-FG02-93ER40756. 
\end{acknowledgments}

\end{document}